\documentclass{article}
\usepackage[final,sglblindworkshop]{neurips_2026}
\workshoptitle{AI for Chip Design}
\usepackage[utf8]{inputenc}
\usepackage[T1]{fontenc}
\usepackage{amsmath}
\usepackage{amssymb}
\usepackage{amsfonts}
\usepackage{booktabs}
\usepackage{graphicx}
\usepackage{subcaption}
\usepackage{array}
\usepackage{multirow}
\usepackage{multicol}
\usepackage{makecell}
\usepackage{tabularx}
\usepackage{enumitem}
\usepackage{xcolor}
\usepackage{hyperref}
\usepackage{url}
\usepackage{nicefrac}
\usepackage{microtype}
\usepackage{float}
\hypersetup{hidelinks}
\setcitestyle{numbers,square}

\newcommand{\cRed}[1]{\textbf{\textcolor{red}{#1}}}
\newcommand{\cBlue}[1]{\textbf{\textcolor{blue}{#1}}}
\newcommand{\Description}[1]{}

\begin{document}

\title{VLM-CAD: VLM-Optimized Collaborative Agent Design Workflow for Analog Circuit Sizing}
\author{%
  \begin{minipage}[t]{0.46\textwidth}\centering
    Guanyuan Pan \\
    Hangzhou Dianzi University, China
  \end{minipage}
  \And
  \begin{minipage}[t]{0.46\textwidth}\centering
    Shuai Wang\thanks{Corresponding Authors: Shuai Wang (\texttt{shuaiwang.tai@gmail.com}) and Yaqi Wang (\texttt{wangyaqi@hdu.edu.cn}).} \\
    Hangzhou Dianzi University, China
  \end{minipage}
  \AND
  \begin{minipage}[t]{0.46\textwidth}\centering
    Yugui Lin \\
    SCBC, Guangdong University \\
    of Foreign Studies, China
  \end{minipage}
  \And
  \begin{minipage}[t]{0.46\textwidth}\centering
    Tiansheng Zhou \\
    Hangzhou Dianzi University, China
  \end{minipage}
  \AND
  \begin{minipage}[t]{0.46\textwidth}\centering
    Pietro Li\`o \\
    University of Cambridge, \\
    United Kingdom
  \end{minipage}
  \And
  \begin{minipage}[t]{0.46\textwidth}\centering
    Zhenxin Zhao \\
    Hangzhou Dianzi University, China
  \end{minipage}
  \AND
  \begin{minipage}[t]{0.46\textwidth}\centering
    Yaqi Wang\footnotemark[1] \\
    Hangzhou Dianzi University, China
  \end{minipage}
}
\maketitle

\begin{abstract}
Vision Language Models (VLMs) have demonstrated remarkable potential in multimodal reasoning. However, they can have spatial blindness and logical hallucinations when interpreting densely structured engineering content, such as analog circuit schematics. To address these challenges, we propose a Vision Language Model-Optimized Collaborative Agent Design Workflow for Analog Circuit Sizing (VLM-CAD) designed to support step-by-step reasoning over multimodal evidence. VLM-CAD bridges the modality gap by integrating a neuro-symbolic structural parsing module, Image2Net, which transforms raw pixels into explicit topological graphs and structured JSON representations to anchor VLM interpretation in deterministic facts. To ensure the reliability required for engineering decisions, we further propose ExTuRBO, an Explainable Trust Region Bayesian Optimization method. ExTuRBO employs agent-generated semantic seeds to warm-start local searches and uses Automatic Relevance Determination to provide sensitivity evidence for the final design report. Experimental results on 12 sizing tasks covering six circuits and four technology platforms show that VLM-CAD achieves a pooled Strict Pass@1 of 23.3\% and a Relaxed Pass@1 of 91.7\%, while providing sensitivity evidence for final design reports.
\end{abstract}


\begin{figure*}
\centering
  \includegraphics[width=\textwidth]{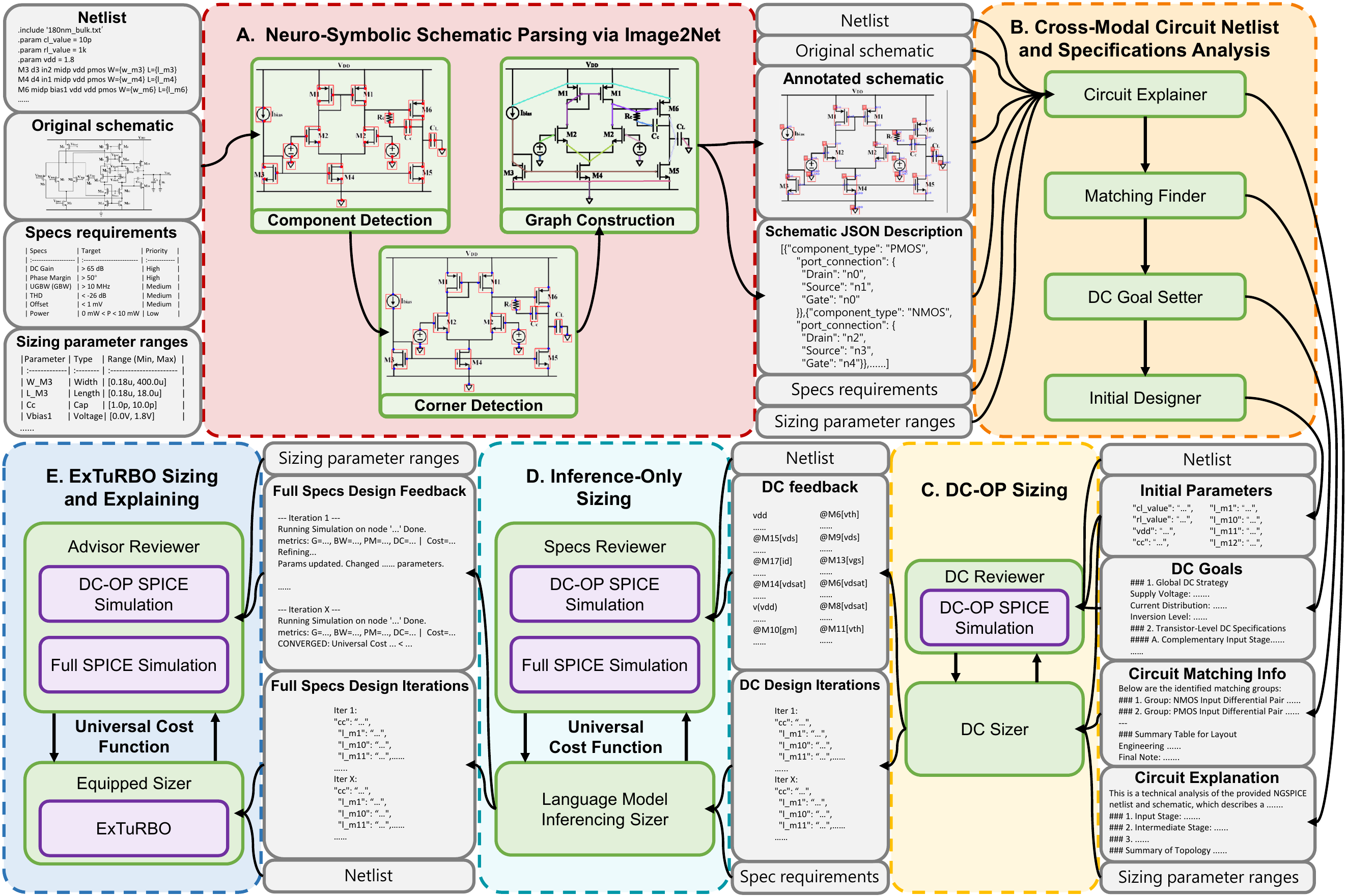}
  \caption{VLM-CAD overview. We include five stages for VLM-CAD: A. Neuro-Symbolic Schematic Parsing via Image2Net (see Sec.~\ref{text:Phase_A}), B. Cross-Modal Circuit Netlist and Specifications Analysis (see Sec.~\ref{text:Phase_B}), C. DC-OP Sizing (see Sec.~\ref{text:Phase_C}), D. Inference-Only Sizing (see Sec.~\ref{text:Phase_D}) and E. ExTuRBO Sizing and Explaining (see Sec.~\ref{text:Phase_E}).}
  \Description{VLM-CAD overview. We include five stages for VLM-CAD: A. Image2Net Interpreting Circuit Schematic (see Sec.~\ref{text:Phase_A}), B. Circuit Netlist and Specifications Analysis (see Sec.~\ref{text:Phase_B}), C. DC-OP Sizing (see Sec.~\ref{text:Phase_C}), D. Inference-Only Sizing (see Sec.~\ref{text:Phase_D}) and E. Sizing and Explaining (see Sec.~\ref{text:Phase_E}).}
  \label{fig:VLM-CAD}
\end{figure*}

\section{Introduction}
Analog and mixed-signal (AMS) integrated circuits are essential components in modern electronic systems. However, the design process, particularly circuit sizing, remains a significant challenge. Analog circuit sizing requires navigating high-dimensional spaces to trade off Power, Performance, and Area to achieve optimal circuit performance. Expert designers use their intuitive understanding of circuit schematics to make decisions. However, existing automatic analog circuit sizing approaches primarily depend on textual netlists, creating a cognitive gap as netlists provide only textual descriptions of components and their connections, lacking the global topological context and spatial cues that schematics offer for high-level functional reasoning~\cite{10.1145_3551901.3556486}.

Vision Language Models (VLMs) have recently demonstrated exceptional capabilities for multimodal reasoning, with the potential to bridge between visual schematics and logical design decisions. However, applying VLMs to automatic analog circuit sizing exposes two critical challenges:

\begin{itemize}[leftmargin=*]
    \item \textbf{Spatial Blindness in Dense Visual Topologies:} Circuit schematics are intersection-heavy and structurally dense. State-of-the-art (SOTA) VLMs exhibit a critical spatial blindness in such contexts, struggling with low-level tasks like identifying line junctions or counting parallel components~\cite{Rahmanzadehgervi_2024_ACCV, vo2026vision}. This failure to accurately map visual geometric relationships to logical functional blocks leads to topological hallucinations, making pure neural-based language model reasoning unreliable for complex circuit analysis~\cite{11262742}.
    \item \textbf{Lack of Quantified Explainability:} Industrial-grade design requires rigorous justification for every transistor size. While using language models for automatic analog circuit sizing offers some level of explainability \cite{ahmadzadeh2025anaflowagenticllmbasedworkflow}, this is questionable due to their tendency to hallucinate, often lacking a ground-truth link to mathematical or physical reality~\cite{bang2025hallulensllmhallucinationbenchmark, 10691781, skelic2025circuitbenchmarkcircuitinterpretation}. Without a quantified sensitivity analysis that clarifies the relationship between spatial visual features and performance metrics, automatic analog circuit sizing methods remain experimental curiosities rather than practical tools for production.
\end{itemize}

To overcome these challenges:

\begin{enumerate}
    \item We propose \textbf{VLM-CAD}, a VLM-optimized collaborative agent workflow for robust multimodal reasoning in analog circuit sizing. VLM-CAD follows the step-by-step multimodal reasoning process used by expert designers. By integrating visual topology with textual specifications, VLM-CAD deduces latent constraints and performs hierarchical reasoning from DC biasing to full-spec optimization.
    \item We integrate \textbf{Image2Net}, a neuro-symbolic structural parsing framework. Image2Net addresses VLM spatial blindness by extracting structured representations from schematics and providing a robust visual memory that anchors the model's reasoning in precise topological facts.
    \item We propose \textbf{ExTuRBO}, an Explainable Trust Region Bayesian Optimization method. ExTuRBO collaboratively warm-starts local search to accelerate convergence and utilizes Automatic Relevance Determination (ARD) lengthscales to provide simulation-grounded, dual-granularity sensitivity evidence for the final design report.
\end{enumerate}

We evaluate VLM-CAD on 12 sizing tasks across six circuits and four technology platforms, improving pooled Strict Pass@1 from 3.3\% to 23.3\% relative to the baseline, with 91.7\% Relaxed Pass@1. We demonstrate VLM-CAD in Figure~\ref{fig:VLM-CAD} and compare it with SOTA methods in Table~\ref{tab:comparison}.

\begin{table}
\centering
\caption{Comparison of VLM-CAD features with SOTA methods. VLM-CAD incorporates schematic and netlist multimodality, LLM with ARD for quantified explainability, and LLM with ExTuRBO for Collaborative Semantic Warm-Start initialization strategy.}
\label{tab:comparison}
\footnotesize
\setlength{\tabcolsep}{1pt}
\renewcommand{\arraystretch}{1.1}
\begin{tabular*}{\linewidth}{@{\extracolsep{\fill}}>{\raggedright\arraybackslash}p{2.55cm}>{\raggedright\arraybackslash}p{2.2cm}>{\raggedright\arraybackslash}p{2.2cm}>{\raggedright\arraybackslash}p{2.5cm}>{\raggedright\arraybackslash}p{4.0cm}@{}}
\toprule
\textbf{Framework} & \makecell[l]{\textbf{Reasoning}\\\textbf{Modality}} & \makecell[l]{\textbf{Topology}\\\textbf{Awareness}} & \textbf{Explainability} & \makecell[l]{\textbf{Initialization}\\\textbf{Strategy}} \\
\midrule
\makecell[l]{LLMACD~\cite{11101244}} & Netlist (Text) & Netlist Inference & LLM CoT (Qualitative) & LLM \\
\makecell[l]{AnaFlow~\cite{ahmadzadeh2025anaflowagenticllmbasedworkflow}} & Netlist (Text) & Netlist Inference & LLM (Qualitative) & \makecell[l]{LLM + Black-box Optimizer} \\
\makecell[l]{EESizer~\cite{Liu_2025}} & Netlist (Text) & Netlist Inference & LLM CoT (Qualitative) & LLM \\
\midrule
\textbf{VLM-CAD (Ours)} & \textbf{Schematic + Netlist} & \textbf{Schematic + Netlist Inference} & \textbf{LLM + ARD (Quantified)} & \textbf{LLM + ExTuRBO (Collaborative Semantic Warm-Start)} \\
\bottomrule
\end{tabular*}
\end{table}

\section{Related Work}

\subsection{Automatic Circuit Sizing: From ML to Agents}
Automatic analog circuit sizing has evolved from traditional optimization to modern agentic workflows. Early efforts primarily used Machine Learning (ML) methods, such as Bayesian Optimization and Reinforcement Learning, to explore high-dimensional design spaces~\cite{electronics14173541}. However, these black-box approaches relied exclusively on textual netlists, lacking the high-level functional hierarchies, such as symmetric differential pairs or feedback loops, that are immediately apparent in a visual schematic. Therefore, they unthinkingly explore the design space without understanding the structural design intent, often leading to simulation-intensive, unintuitive search processes \cite{SavidisIoannis2025EMTf, 1665031}.

Recently, Large Language Models (LLMs) have demonstrated their capabilities to optimize analog circuits. LEDRO~\cite{11106097} utilized LLMs to prune high-dimensional search spaces through design space reduction, while LLMACD~\cite{11101244} incorporated circuit behavior knowledge into prompt-based sizing. To enhance autonomous decision-making, multi-agent frameworks like AmpAgent~\cite{liu2024ampagentllmbasedmultiagentmultistage} and AnaFlow~\cite{ahmadzadeh2025anaflowagenticllmbasedworkflow} replicated expert design hierarchies and decomposed complex design tasks into subtasks, such as topology selection and parameter refinement. However, textual LLMs struggle to accurately reason about AMS circuit topologies~\cite{skelic2025circuitbenchmarkcircuitinterpretation}, leading to sub-optimal trade-offs. Furthermore, the inherent lack of mathematical precision in LLMs~\cite{Terpstra2024EmpoweringAnalog} and their susceptibility to hallucination~\cite{bang2025hallulensllmhallucinationbenchmark} underscore the critical need for multimodal alignment: integrating visual topological context with textual specifications and numerical simulation feedback to ensure logical consistency.

\subsection{Schematic Understanding and Spatial Reasoning in VLMs}
Circuit schematics visually depict the structure and interconnections of a circuit, providing designers with an intuitive reference for understanding how it operates~\cite{10.1145_3551901.3556486}. However, automatic circuit schematics interpretation represents a challenging domain for dense diagrammatic reasoning. While VLMs have shown promise in general image captioning, their application to structured engineering documents is still in its infancy. MAPS~\cite{zhu2025maps} finetunes a VLM to convert circuit schematics into SPICE netlists, while CURVLM~\cite{zhaocurvlm} finetunes a VLM for analytical reasoning on digital circuit schematics.
Despite these advancements, purely neural-based VLM architectures face fundamental bottlenecks in high-precision spatial tasks. Even SOTA VLMs exhibited persistent spatial blindness, struggling with reliable distance estimation and direction comprehension in dense layouts~\cite{tang-etal-2025-sparkle}, leading to topological hallucinations, where the model misinterpreted line intersections or miscounts recurring geometric shapes~\cite{Rahmanzadehgervi_2024_ACCV, charthal, vo2026vision}. In analog design, even a single pixel of connectivity error can invalidate the entire physical logic. Therefore, purely neural-based reasoning remains unreliable~\cite{11262742, 10691781}.

\section{VLM-CAD Agentic Workflow}

\subsection{Neuro-Symbolic Schematic Parsing via Image2Net}
\label{text:Phase_A}

Despite excellent performance on object detection and general multimodal reasoning tasks, VLMs inherently struggle with spatial blindness in densely structured diagrams, often failing to correctly count intersections or map pixel-level geometric relationships to logical functional blocks, leading to topological hallucinations. To address this, we utilize a neuro-symbolic schematic parsing approach. Specifically, we employ Image2Net~\cite{edachallenge2024} to transform raw pixel data into an explicit symbolic representation, thereby anchoring the VLM's high-level reasoning in deterministic topological facts.

\begin{itemize}[leftmargin=*]
    \item \textbf{Component Detection:} We finetune a YOLOv8-Pose model~\cite{10533619} on the train dataset from~\cite{edachallenge2024}, which contains 2268 noise-free circuit schematics, and employ it as the neural front-end to identify component bounding boxes and predict terminal pose coordinates. This neural layer handles the visual variety of schematic symbols, translating visual primitives into identifiable objects.
    \item \textbf{Corner Detection:} To capture the wiring topology, we apply deterministic feature detection algorithms followed by Density-Based Spatial Clustering of Applications with Noise clustering~\cite{3001460.3001507}. By merging duplicate detections into precise geometric nodes, we mitigate the noise inherent in neural-only detection.
    \item \textbf{Graph Construction:} We transform pixel-level visual continuity into an explicit connectivity graph through a hybrid algorithm: vectorized array slicing validates orthogonal connections, while Bresenham's algorithm traces diagonal paths. By clustering physically connected nodes into distinct electrical nets, we provide a symbolic abstraction of the visual topology.
    \item \textbf{Annotated Schematic and JSON Description Generation:} Finally, we map the parsed graph to a structured symbolic JSON description. This description serves as an explicit, topology-grounded visual memory for the VLM, helping the subsequent agentic design workflow maintain consistent access to the parsed schematic structure. Additionally, we create an annotated schematic that color-codes the distinct electrical connections for verification.
\end{itemize}

We present the details of this phase in Figure~\ref{fig:VLM-CAD}.A.

\subsection{Cross-Modal Circuit Netlist and Specifications Analysis}
\label{text:Phase_B}
After annotating the schematic, we analyze the sizing problem using four specialist agents:
\begin{itemize}[leftmargin=*]
    \item \textbf{Circuit Explainer}: ingests the schematic, netlist, and specifications to identify analog sub-blocks, deduce signal flow and feedback, and produce a functional circuit description.
    \item \textbf{Matching Finder}: identifies transistor clusters requiring symmetry and emits parameter-level equality constraints. It also infers layout implications such as interdigitation needs.
    \item \textbf{DC Goal Setter}: assigns each transistor quantitative operating targets: overdrive voltage $V_{\mathrm{ov}}$, drain-source voltage $V_{\mathrm{DS}}$, and current density. It also defines a global biasing plan that reserves sufficient headroom for signal swing and includes the start-up conditions required for convergence.
    \item \textbf{Initial Designer}: parses the netlist to extract all mandatory design variables and populates each key with a conservative executable value aligned with the DC goals, mitigating parameter hallucination and omission.
\end{itemize}

\begin{figure}
\centering
\begin{minipage}{0.60\linewidth}
    \centering
    \includegraphics[width=\linewidth]{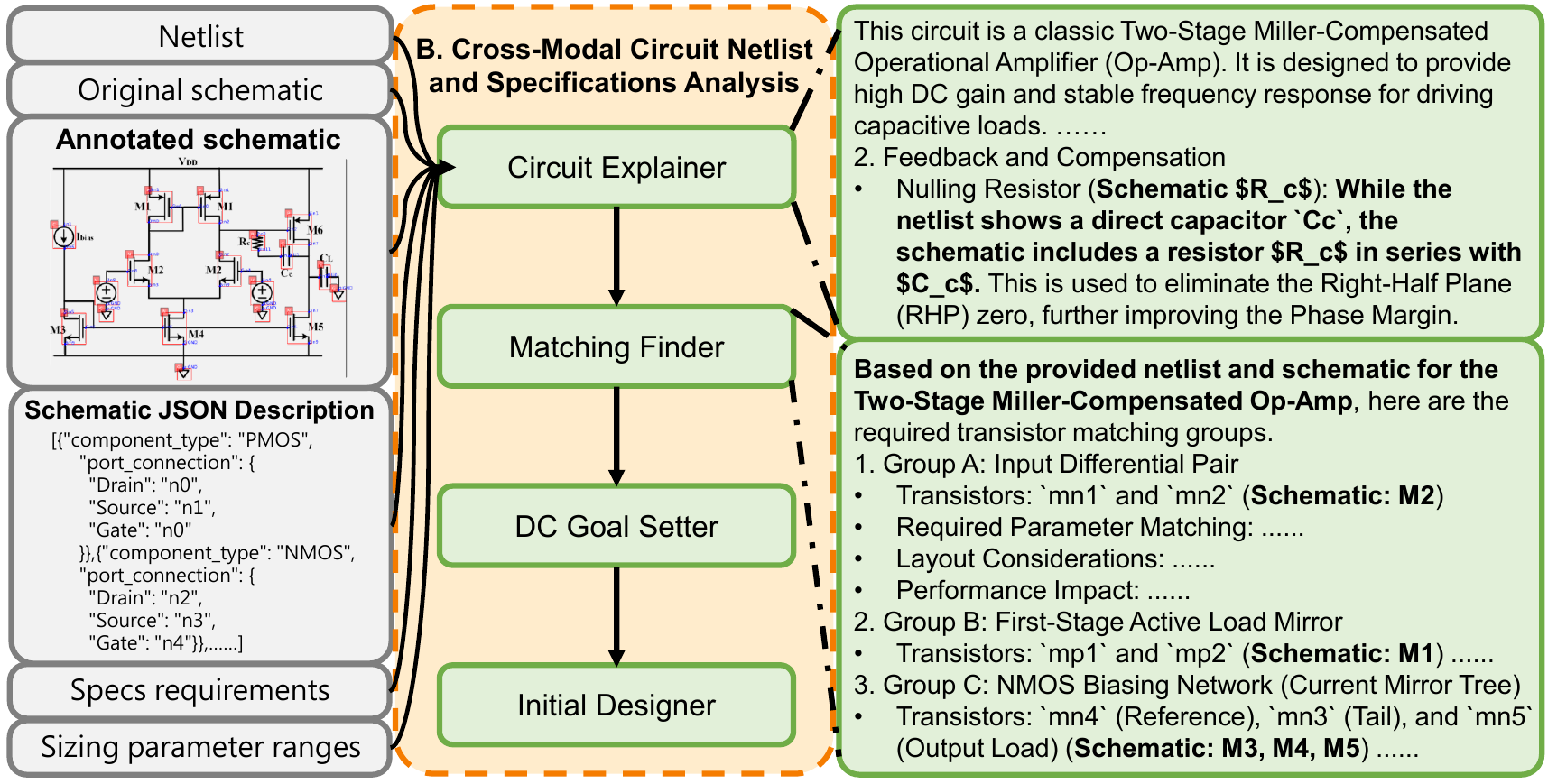}
    \captionof{figure}{Circuit netlist and specifications analysis in VLM-CAD. The VLM-based circuit explainer and matching finder utilize both the circuit schematic and netlist for multimodal reasoning.}
    \Description{Circuit netlist and specifications analysis in VLM-CAD. The VLM-based circuit explainer and matching finder utilize both the circuit schematic and netlist for multimodal reasoning.}
    \label{fig:phase1}
\end{minipage}%
\hfill%
\begin{minipage}{0.38\linewidth}
    \centering
    \scriptsize
    \captionof{table}{Comparative analysis of multimodal reasoning performance between VLMs and Image2Net-grounded parsing.}
    \label{tab:schematic_detection}
    \setlength{\tabcolsep}{1pt}
    \renewcommand{\arraystretch}{1.25}
    \begin{tabular}{@{}>{\raggedright\arraybackslash}m{1.3cm}>{\centering\arraybackslash}m{0.72cm}>{\centering\arraybackslash}m{1.3cm}>{\centering\arraybackslash}m{2.2cm}@{}}
    \toprule
    \shortstack{\textbf{Method}} & \shortstack{\textbf{Non-}\\\textbf{Empty}\\\textbf{Output}} & \shortstack{\textbf{Functional}\\\textbf{Accuracy}\\$F(\%)$} & \shortstack{\textbf{Connectivity}\\\textbf{Score }$S_k$} \\
    \midrule
    \shortstack[l]{Gemini 3\\Flash Preview} & 122 & 95.90 & 0.6600 $\pm$ 0.1652 \\ 
    GPT-5.2 & 119 & 49.18 & 0.6096 $\pm$ 0.1278 \\ 
    \shortstack[l]{Gemma-4\\31B-it} & 122 & 52.46 & 0.5918 $\pm$ 0.1202 \\ 
    \shortstack[l]{Qwen3\\VL-235B\\A22B-Instruct} & 119 & 58.20 & 0.5751 $\pm$ 0.1091 \\ 
    \midrule
    \textbf{Image2Net} & \textbf{122} & \textbf{68.85} & \textbf{0.8121 $\pm$ 0.2116} \\ 
    \bottomrule
    \end{tabular}
\end{minipage}
\end{figure}

We present the details of this phase in Figure~\ref{fig:VLM-CAD}.B~and~\ref{fig:phase1}.

\subsection{DC-OP Sizing}
\label{text:Phase_C}
VLM-CAD first establishes a DC-biased solution before costly performance simulations:
\begin{itemize}[leftmargin=*]
    \item \textbf{DC Reviewer}: runs a fast DC operating-point simulation at each iteration, compares node voltages and device operating points against the DC goals from previous analysis, and computes a discrepancy count as the loop stopping criterion.
    \item \textbf{DC Sizer}: updates parameters from the discrepancy report using analog design heuristics while preserving static testbench values, matching constraints, and physical bounds across iterations.
\end{itemize}

We present the details of this phase in Figure~\ref{fig:VLM-CAD}.C.

\subsection{Inference-Only Sizing}
\label{text:Phase_D}
After DC-OP sizing, VLM-CAD optimizes all specifications through full simulations, relying solely on the inherent knowledge of analog circuit theory provided by language models without any external numerical optimizers.

We convert the multi-objective sizing task into a scalar \textbf{Universal Cost Function} $J(\mathbf{x})$ shared by the language model agents and the numerical optimizer. It has two modes: \textbf{Feasibility}, which prioritizes satisfying all specifications except power, and \textbf{Optimization}, which minimizes power. For a design $\mathbf{x}$,
\begin{equation}
\label{eq:cost_func}
J(\mathbf{x}) = \frac{P_{\text{meas}}(\mathbf{x})}{P_{\text{max}}} + \sum_{i \in \mathcal{S}} w_i \cdot \mathcal{V}_i(y_i(\mathbf{x}), T_i) + \mathcal{P}_{\text{sanity}}.
\end{equation}
Here, $P_{\text{meas}}$ is the simulated power and $P_{\text{max}}$ its limit; $\mathcal{S}$ is the set of performance metrics, split into lower-bound metrics $\mathcal{S}_{\text{LB}}$ such as gain and phase margin, and upper-bound metrics $\mathcal{S}_{\text{UB}}$ such as THD and offset; $w_i$ is the penalty weight of metric $i$; and $\mathcal{P}_{\text{sanity}}$ is a large constant penalty applied only when basic functionality is lost, such as non-convergence or near-zero gain. The violation term $\mathcal{V}_i$ enforces the direction of each constraint:
\begin{equation}
\label{eq:violation}
 \mathcal{V}_i(y_i(\mathbf{x}), T_i) = 
 \begin{cases} 
 \max(0, T_i - y_i(\mathbf{x})), & i \in \mathcal{S}_{\text{LB}} \\
 \max(0, y_i(\mathbf{x}) - T_i), & i \in \mathcal{S}_{\text{UB}}.
 \end{cases}
\end{equation}
$J(\mathbf{x})$ separates two operating regimes rather than encoding feasibility by construction: above the regime boundary the sizer prioritizes driving every specification except power toward its target, and within the low-cost regime it shifts effort to minimizing power under periodic re-verification of each specification against its target. In our runs, the boundary was set at $J(\mathbf{x}) \le 1$, and a sizing loop terminated when either $J(\mathbf{x}) < 0.5$ or every hard specification of the task was met by the simulated response.

We present the details of this phase in Figure~\ref{fig:VLM-CAD}.D:

\begin{itemize}[leftmargin=*]
    \item \textbf{Specs Reviewer}: applies a Skip-on-Fail evaluation: it verifies the DC operating point first and halts early if biasing fails, then runs full DC/AC simulations and evaluates $J(\mathbf{x})$ via \eqref{eq:cost_func}.
    \item \textbf{Language Model Inferencing Sizer}: updates parameters from a history-aware context to avoid oscillation, alternating between satisfying constraints and minimizing power. If a proposal repeats the previous one, a dead-loop detector applies a small random perturbation to transistor widths to break the stagnation.
\end{itemize}

\subsection{ExTuRBO Sizing and Explaining}
\label{text:Phase_E}
When inference-only sizing reaches its iteration limit or convergence plateau, VLM-CAD escalates to \textbf{ExTuRBO}, an Explainable Trust Region Bayesian Optimization method that provides explainability for the final design report in the form of sensitivity evidence, through two features:

\begin{itemize}[leftmargin=*]
    \item \textbf{Collaborative Semantic Warm-Start}: Standard TuRBO initializes with Latin Hypercube Sampling (LHS)~\cite{NEURIPS2019_6c990b7a} over the full design space, wasting simulation budget on non-functional regions. ExTuRBO instead seeds its search from the inference-only phase: the Advisor selects up to three distinct Phase-D candidates from the low-cost regime ($J(\mathbf{x}) < 1$), and each parallel worker starts from one candidate with locally contracted bounds around it. Relative to a global search over $\mathcal{X}_{\text{global}}\subset\mathbb{R}^D$, this prunes the search to a local subspace $\mathcal{X}_{\text{local}}$ with span ratio $r<1$, so the search volume shrinks exponentially in the dimension:
    \begin{equation}
    \label{eq:volume}
    \frac{V_{\text{local}}}{V_{\text{global}}} \approx \prod_{d=1}^{D} r_d \approx r^D.
    \end{equation}
    For example, in a 48-dimensional problem with $r=0.4$, we reduce the volume by a factor of $\sim 10^{19}$, substantially reducing global exploration and focusing optimization on high-quality regions identified by the agents.

    \item \textbf{Grounding via Dual-Granularity Sensitivity}: After optimization, ExTuRBO fits two ARD Gaussian processes to produce simulation-grounded sensitivity evidence for the final design report: \textbf{Global Sensitivity}, fit on all evaluated designs, identifies survival parameters most associated with basic circuit feasibility, and \textbf{Elite Sensitivity}, fit on the top $15\%$ of designs, identifies tuning parameters most associated with high performance. We derive feature importance from the ARD lengthscales as $S_d \propto 1/\ell_d$: the smaller $\ell_d$, the more strongly the modeled cost depends on parameter $d$.
\end{itemize}

We present the details of this phase in Figure~\ref{fig:VLM-CAD}.E:

\begin{itemize}[leftmargin=*]
    \item \textbf{Advisor Reviewer}: analyzes the inference-only history, deduplicates and selects the best distinct low-cost candidates as seeds, and configures ExTuRBO's local search bounds.
    \item \textbf{Equipped Sizer}: runs the parallel ExTuRBO workers on these seeds and converts the Global and Elite sensitivity results into a design sign-off report that separates stability-critical parameters from performance-tuning parameters.
\end{itemize}%

\section{Experiments}
\label{sec:experiments}
\subsection{Multimodal Structural Reasoning Performance}
To assess Image2Net-grounded structural parsing, we evaluate schematic parsing on 122 schematics that contain significant visual noise from~\cite{edachallenge2024}, which are more visually degraded than those used in VLM-CAD (see Figure~\ref{fig:circuit_topologies}). We compare Image2Net with four general-purpose VLMs: Gemini 3 Flash Preview~\cite{google2025gemini3flash}, GPT-5.2~\cite{openai2025gpt52}, Gemma-4-31B-it~\cite{gemmateam2026gemma4technicalreport}, and Qwen3-VL-235B-A22B-Instruct~\cite{bai2025qwen3vltechnicalreport}. Our comparison focuses on functional accuracy $F$, the percentage of schematics whose predicted circuit-type label matches the reference, and the connectivity score $S_k=1/\log_{10}(10+K)$, where $K$ is the graph edit cost between the predicted and reference connectivity graphs. The complete scoring protocol is provided in Appendix~\ref{app:structural_protocol}.

Results show that the VLMs exhibit clear spatial blindness: functional accuracy varies widely, and connectivity scores are consistently weaker than Image2Net's. Common errors include misreading line crossings as junctions and miscounting parallel transistors, and both GPT-5.2 and Qwen3-VL-235B-A22B-Instruct return blank or non-terminating outputs on some schematics.

However, Image2Net is a specialized neuro-symbolic parser that we train on 2268 schematics and remains imperfect under novel noise. In contrast, VLMs offer zero-shot semantic reasoning: Gemini 3 Flash Preview attains 95.90\% functional accuracy, above Image2Net's 68.85\%. VLM-CAD therefore offloads deterministic netlist recognition to Image2Net while the VLM performs high-level circuit reasoning, grounding the agent workflow in explicit connectivity.

\subsection{Sizing Results}
\begin{figure}[t]
\centering
\begin{subfigure}[b]{0.48\textwidth}
    \centering
    \includegraphics[width=\textwidth]{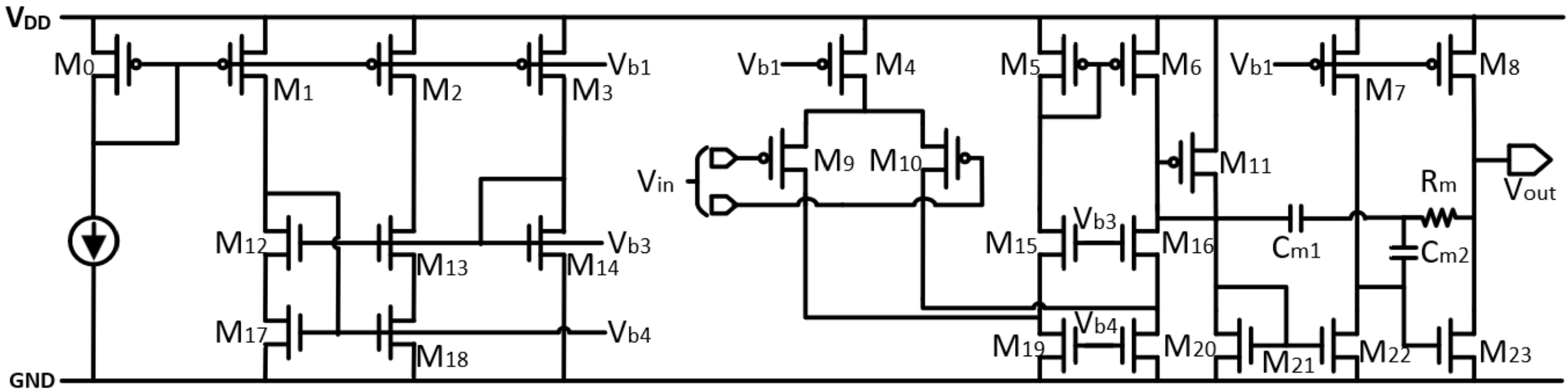}
    \caption{NMCNR.}
\end{subfigure}
\hfill
\begin{subfigure}[b]{0.48\textwidth}
    \centering
    \includegraphics[width=\textwidth]{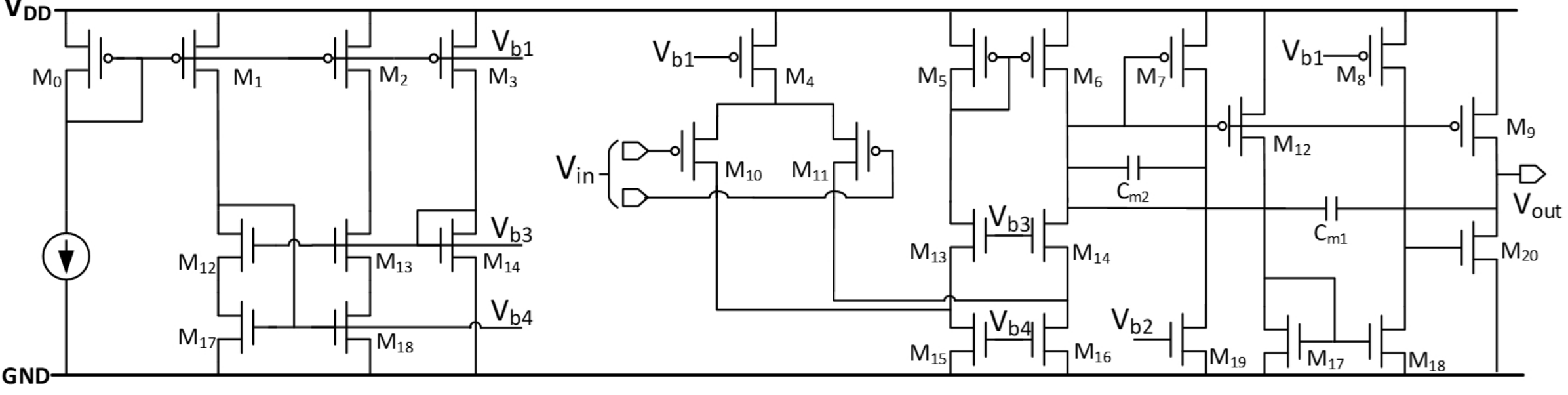}
    \caption{DFCFC.}
\end{subfigure}
\par\vspace{0.6em}

\begin{subfigure}[t]{0.24304\textwidth}%
    \vspace{0pt}
    \centering
    \includegraphics[width=\linewidth]{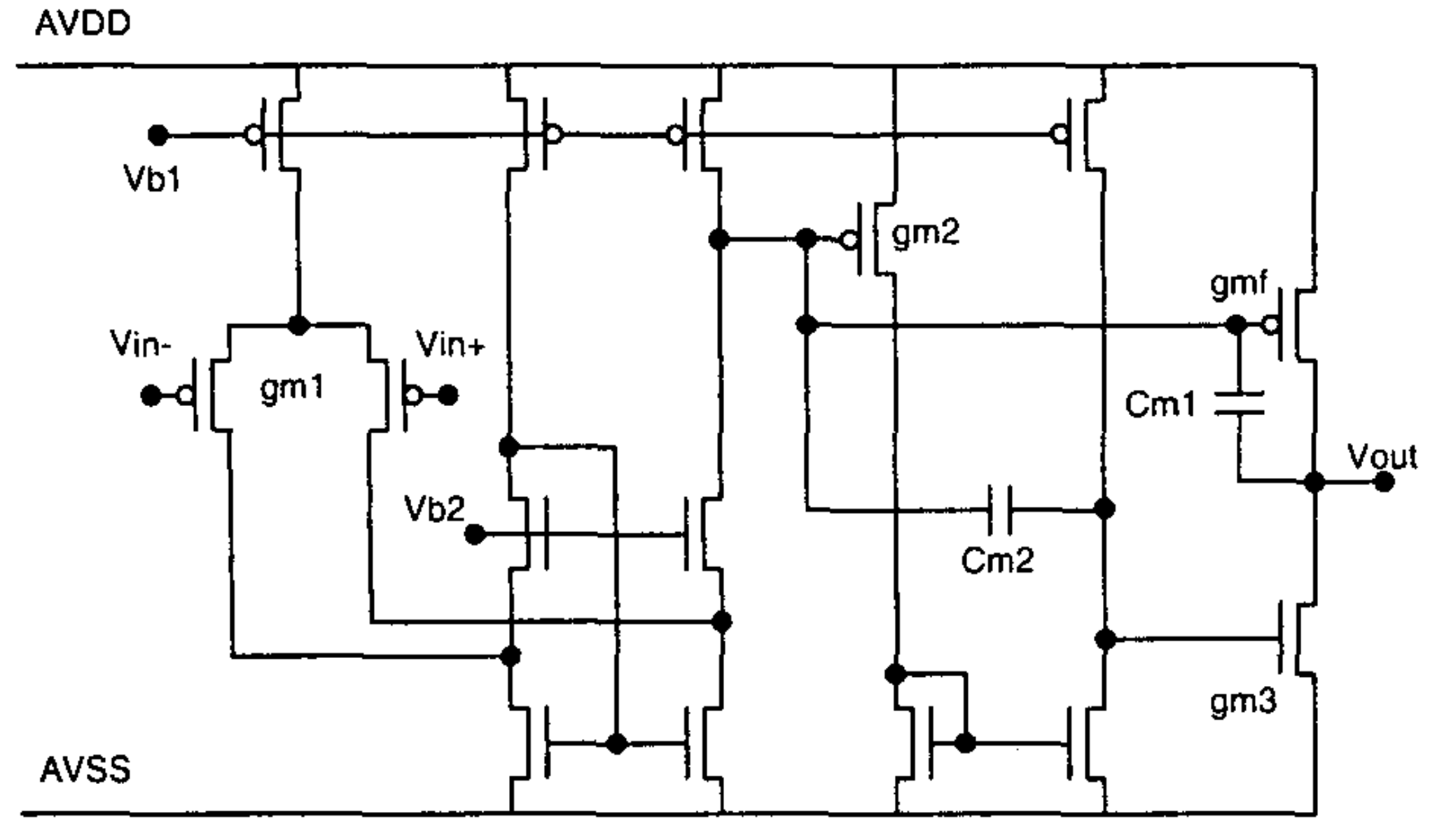}
    \caption{PFC.}
\end{subfigure}%
\begin{subfigure}[t]{0.18865\textwidth}%
    \vspace{0pt}
    \centering
    \includegraphics[width=\linewidth]{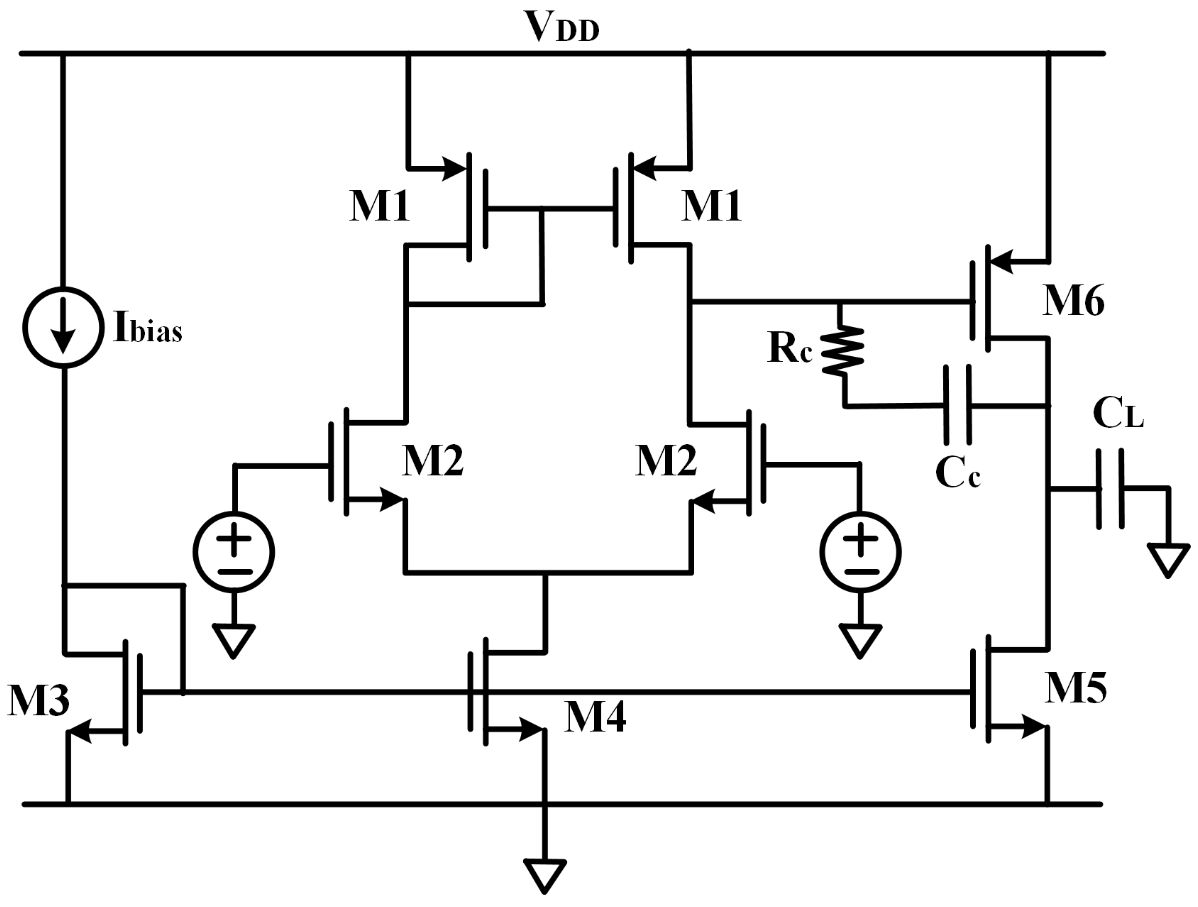}
    \caption{Miller Op-Amp.}
\end{subfigure}%
\begin{subfigure}[t]{0.24121\textwidth}%
    \vspace{0pt}
    \centering
    \includegraphics[width=\linewidth]{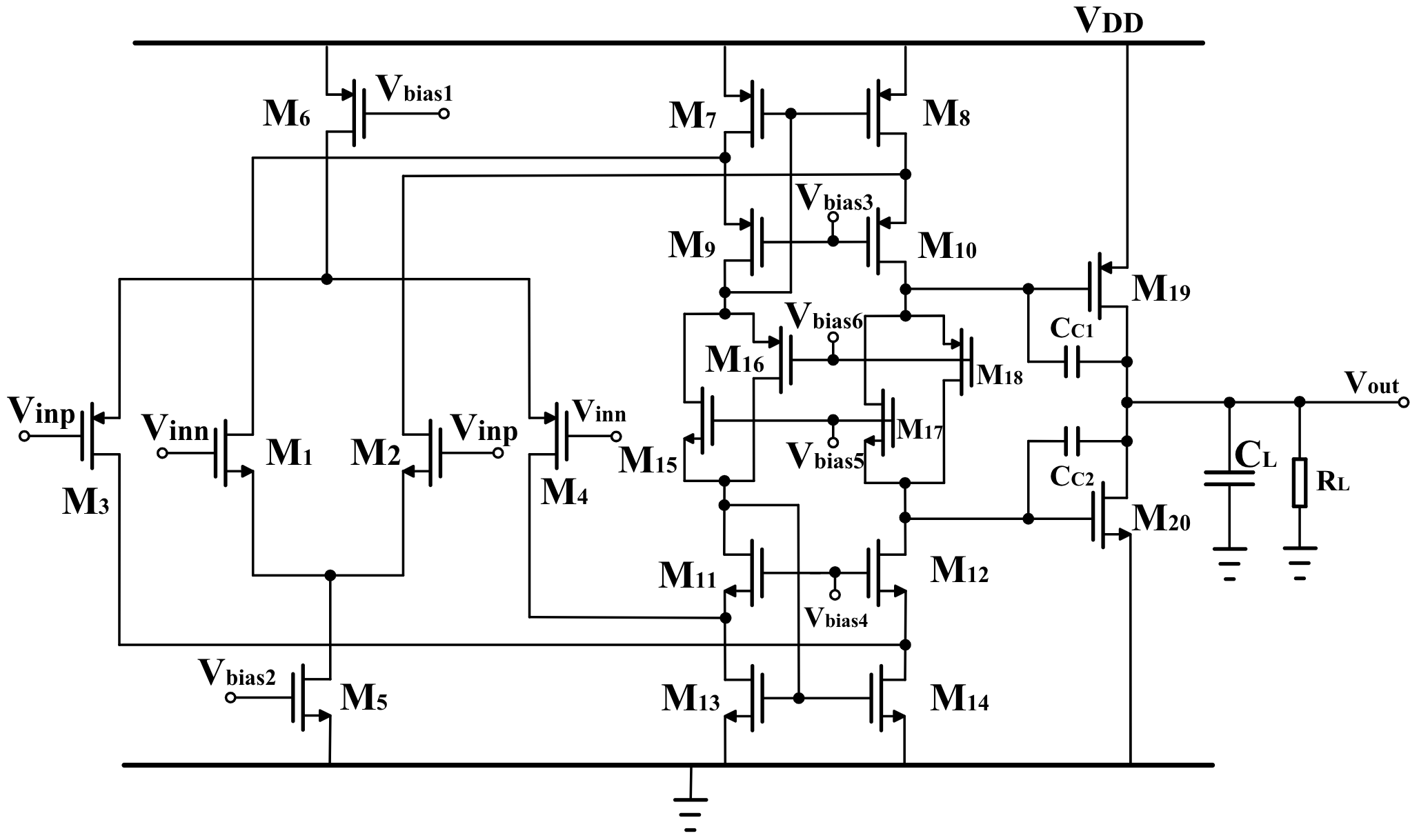}
    \caption{Class-AB Op-Amp.}
\end{subfigure}%
\begin{subfigure}[t]{0.32710\textwidth}%
    \vspace{0pt}
    \centering
    \includegraphics[width=\linewidth]{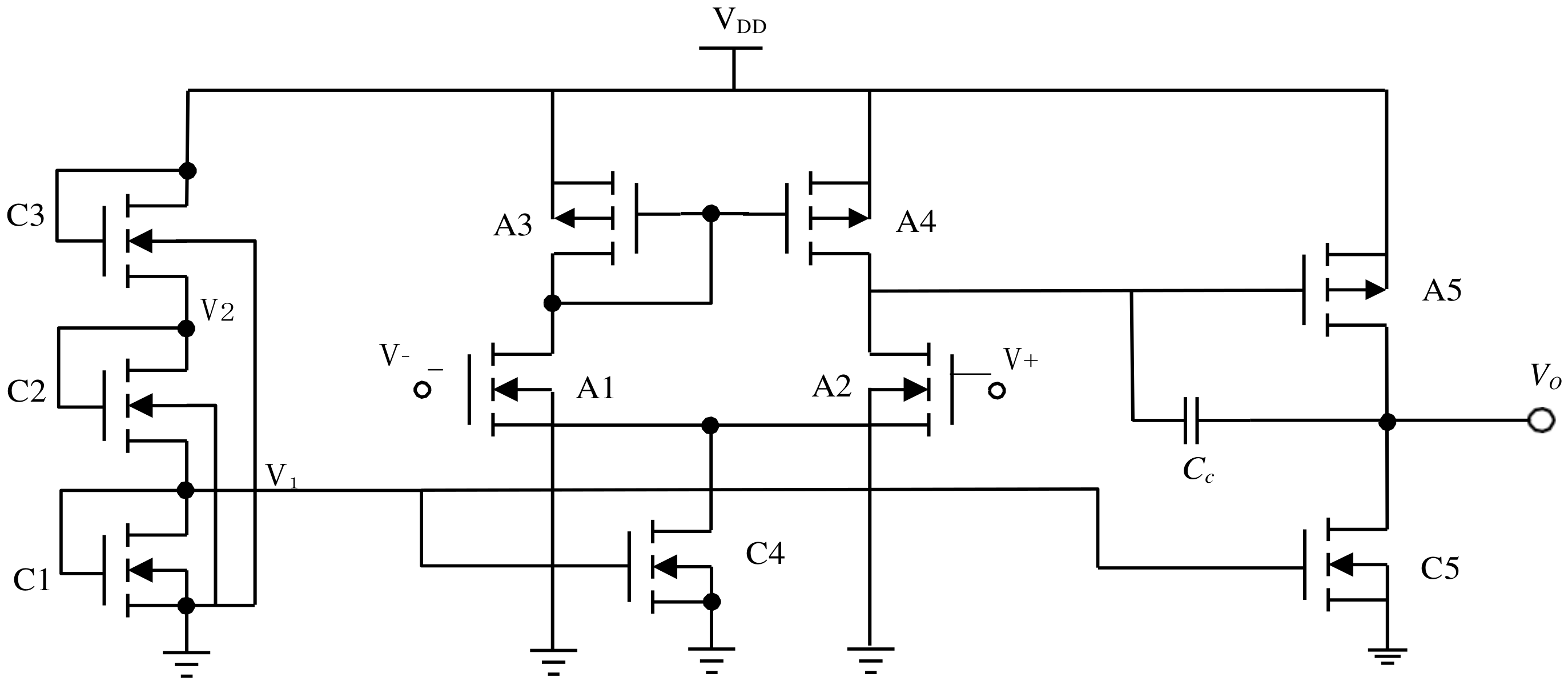}
    \caption{Discrete Op-Amp.}
\end{subfigure}

\caption{Schematic topologies of the six analog circuits we use: (a) a nested Miller-compensated amplifier with a nulling resistor~\cite{11126341,948432}, (b) a damping-factor-control frequency-compensation amplifier~\cite{11126341,948432}, (c) a two-stage Miller operational amplifier~\cite{11063353}, (d) a positive-feedback-compensated amplifier~\cite{11126341,1012833}, (e) a complementary-input class-AB operational amplifier~\cite{Liu_2025}, and (f) a two-stage discrete-MOSFET operational amplifier.}
\label{fig:circuit_topologies}
\end{figure}

We evaluate VLM-CAD on six circuit topologies, shown in Figure~\ref{fig:circuit_topologies}, under two specification sets per circuit. The study spans discrete device models, PTM 180nm and 45nm models~\cite{cao2011predictive}, and the SKY130 technology platform~\cite{edwards2020google}.

\begin{table}[H]
\centering

\caption{Benchmark circuits, technology platforms, and target specifications.
BSIM PTM stands for Berkeley Short-channel IGFET Model Predictive Technology Models~\cite{cao2011predictive}.
SKY130 stands for SkyWater SKY130 PDK~\cite{edwards2020google}.
``---'' indicates that the metric is not measurable for the given circuit.
For readability, THD is reported as a percentage using
$\mathrm{THD}(\%)=100\times10^{\mathrm{THD(dB)}/20}$.}
\label{tab:sizing_circuits}
\centering
\footnotesize
\setlength{\tabcolsep}{2pt}
\renewcommand{\arraystretch}{1.15}
\begin{tabularx}{\textwidth}{@{}>{\raggedright\arraybackslash}l>{\raggedright\arraybackslash}l c *{6}{>{\centering\arraybackslash}X}@{}}
\toprule
\textbf{Circuit} & \textbf{Technology} & \textbf{Specs} &
\shortstack{\textbf{Gain}\\\textbf{(dB)}} & \shortstack{\textbf{PM}\\\textbf{($^\circ$)}} & \shortstack{\textbf{UGBW}\\\textbf{(MHz)}} & \shortstack{\textbf{THD}\\\textbf{(\%)}} & \shortstack{\textbf{Offset}\\\textbf{(mV)}} & \shortstack{\textbf{Power}\\\textbf{(mW)}} \\
\midrule
\multirow{2}{*}{NMCNR} & \multirow{2}{*}{SKY130} & SpecsA & $>90$ & $>55$ & $>2.00$ & $<0.316\%$ & $<1.0$ & $<1.00$ \\
 & & SpecsB & $>80$ & $>60$ & $>1.00$ & $<0.562\%$ & $<1.0$ & $<0.15$ \\
\addlinespace
\multirow{2}{*}{DFCFC} & \multirow{2}{*}{SKY130} & SpecsA & $>90$ & $>50$ & $>3.00$ & $<0.100\%$ & $<0.5$ & $<2.00$ \\
 & & SpecsB & $>80$ & $>45$ & $>5.00$ & $<0.316\%$ & $<1.0$ & $<4.00$ \\
\addlinespace
\multirow{2}{*}{PFC} & \multirow{2}{*}{SKY130} & SpecsA & $>80$ & $>50$ & $>5.00$ & $<0.316\%$ & $<1.0$ & $<1.50$ \\
 & & SpecsB & $>70$ & $>45$ & $>20.00$ & $<1.00\%$ & $<1.0$ & $<1.50$ \\
\addlinespace
\multirow{2}{*}{Miller Op-Amp} & \multirow{2}{*}{\shortstack[l]{PTM\\45\,nm}} & SpecsA & $>54$ & $>60$ & $>1.00$ & $<0.100\%$ & $<5.0$ & $<0.15$ \\
 & & SpecsB & $>48$ & $>50$ & $>18.00$ & $<1.26\%$ & $<8.0$ & $<0.40$ \\
\addlinespace
\multirow{2}{*}{Class-AB Op-Amp} & \multirow{2}{*}{\shortstack[l]{PTM\\180\,nm}} & SpecsA & $>65$ & $>65$ & $>1.00$ & $<5.01\%$ & $<0.5$ & $<0.60$ \\
 & & SpecsB & $>65$ & $>50$ & $>10.00$ & $<5.01\%$ & $<1.0$ & $<10.00$ \\
\addlinespace
\multirow{2}{*}{Discrete Op-Amp} & \multirow{2}{*}{\shortstack[l]{Discrete\\MOSFET}} & SpecsA & $>90$ & $>60$ & $>0.35$ & $<0.178\%$ & $<0.1$ & --- \\
 & & SpecsB & $>85$ & $>50$ & $>0.80$ & $<0.178\%$ & $<0.1$ & --- \\
\bottomrule
\end{tabularx}
\end{table}

Our baseline is EESizer~\cite{Liu_2025}, an agent that performs analog sizing through CoT inference over the netlist. For comparability under each method's default configuration, all methods utilize Gemma-4-31B-it~\cite{gemmateam2026gemma4technicalreport} as their backbone VLMs, with five independent runs per task. Following~\cite{chen2021evaluatinglargelanguagemodels}, for a task with $c$ successful runs out of $n=5$, we compute
\begin{equation}
\mathrm{Pass@}k = 1 - \frac{\binom{n-c}{k}}{\binom{n}{k}}.
\end{equation}
We count a run as Strict when it satisfies all performance and power constraints, and as Relaxed when the power constraint is disregarded and at most one other specification is violated. We report 95\% confidence intervals for pooled $Pass@1$ and $Pass@3$, obtained by a task-level cluster bootstrap over the twelve tasks.

\begin{table}[t]
\footnotesize
\centering
\caption{$Pass@1$ and $Pass@3$ (\%) of EESizer and VLM-CAD across six circuits and two specifications.
Strict: all metrics meet their targets; Relaxed: at most one metric misses its target with the power requirement disregarded.
\textbf{Bold} indicates the best results. 
The last row reports 95\% confidence intervals of the pooled $Pass@1$ and $Pass@3$, obtained by the same task-level bootstrap.}
\label{tab:sizing_results}
\setlength{\tabcolsep}{3pt}
\begin{tabular*}{\textwidth}{@{\extracolsep{\fill}}llccccccccc@{}}
\toprule
\multirow{3}{*}{\textbf{Circuit}} & \multirow{3}{*}{\textbf{Tech}} & \multirow{3}{*}{\textbf{Spec}} & \multicolumn{4}{c}{\textbf{EESizer}} & \multicolumn{4}{c}{\textbf{VLM-CAD}} \\
\cmidrule(lr){4-7}\cmidrule(lr){8-11}
& & & \multicolumn{2}{c}{\textbf{Strict}} & \multicolumn{2}{c}{\textbf{Relaxed}} & \multicolumn{2}{c}{\textbf{Strict}} & \multicolumn{2}{c}{\textbf{Relaxed}} \\
\cmidrule(lr){4-5}\cmidrule(lr){6-7}\cmidrule(lr){8-9}\cmidrule(lr){10-11}
& & & \textbf{P@1} & \textbf{P@3} & \textbf{P@1} & \textbf{P@3} & \textbf{P@1} & \textbf{P@3} & \textbf{P@1} & \textbf{P@3} \\
\midrule
\multirow{2}{*}{\shortstack[l]{NMCNR}} & \multirow{2}{*}{SKY130} & SpecsA & 0.0 & 0.0 & 0.0 & 0.0 & 0.0 & 0.0 & \textbf{100.0} & \textbf{100.0} \\
 & & SpecsB & 0.0 & 0.0 & 0.0 & 0.0 & 0.0 & 0.0 & \textbf{100.0} & \textbf{100.0} \\ \addlinespace
\multirow{2}{*}{\shortstack[l]{DFCFC}} & \multirow{2}{*}{SKY130} & SpecsA & 0.0 & 0.0 & 0.0 & 0.0 & 0.0 & 0.0 & \textbf{100.0} & \textbf{100.0} \\
 & & SpecsB & 0.0 & 0.0 & 0.0 & 0.0 & \textbf{60.0} & \textbf{100.0} & \textbf{100.0} & \textbf{100.0} \\ \addlinespace
\multirow{2}{*}{\shortstack[l]{PFC}} & \multirow{2}{*}{SKY130} & SpecsA & 0.0 & 0.0 & 0.0 & 0.0 & \textbf{40.0} & \textbf{90.0} & \textbf{80.0} & \textbf{100.0} \\
 & & SpecsB & 0.0 & 0.0 & 0.0 & 0.0 & 0.0 & 0.0 & \textbf{80.0} & \textbf{100.0} \\ \addlinespace
\multirow{2}{*}{\shortstack[l]{Miller\\Op-Amp}} & \multirow{2}{*}{\shortstack[l]{PTM\\45nm}} & SpecsA & 0.0 & 0.0 & 0.0 & 0.0 & 0.0 & 0.0 & \textbf{100.0} & \textbf{100.0} \\
 & & SpecsB & 0.0 & 0.0 & 0.0 & 0.0 & 0.0 & 0.0 & \textbf{40.0} & \textbf{90.0} \\ \addlinespace
\multirow{2}{*}{\shortstack[l]{Class-AB\\Op-Amp}} & \multirow{2}{*}{\shortstack[l]{PTM\\180nm}} & SpecsA & 20.0 & 60.0 & \textbf{100.0} & \textbf{100.0} & \textbf{60.0} & \textbf{100.0} & \textbf{100.0} & \textbf{100.0} \\
 & & SpecsB & 20.0 & 60.0 & 80.0 & \textbf{100.0} & \textbf{40.0} & \textbf{90.0} & \textbf{100.0} & \textbf{100.0} \\ \addlinespace
\multirow{2}{*}{\shortstack[l]{Discrete\\Op-Amp}} & \multirow{2}{*}{\shortstack[l]{Discrete\\MOSFET}} & SpecsA & 0.0 & 0.0 & 0.0 & 0.0 & \textbf{40.0} & \textbf{90.0} & \textbf{100.0} & \textbf{100.0} \\
 & & SpecsB & 0.0 & 0.0 & 0.0 & 0.0 & \textbf{40.0} & \textbf{90.0} & \textbf{100.0} & \textbf{100.0} \\
\midrule
\multicolumn{3}{l}{Average} & 3.3 & 10.0 & 15.0 & 16.7 & \textbf{23.3} & \textbf{46.7} & \textbf{91.7} & \textbf{99.2} \\
\multicolumn{3}{l}{\scriptsize 95\% CI} & \shortstack{\scriptsize [0.0,\\ \scriptsize 8.3]} & \shortstack{\scriptsize [0.0,\\ \scriptsize 25.0]} & \shortstack{\scriptsize [0.0,\\ \scriptsize 36.7]} & \shortstack{\scriptsize [0.0,\\ \scriptsize 41.7]} & \shortstack{\scriptsize [10.0,\\ \scriptsize 36.7]} & \shortstack{\scriptsize [22.5,\\ \scriptsize 71.7]} & \shortstack{\scriptsize [80.0,\\ \scriptsize 100.0]} & \shortstack{\scriptsize [97.5,\\ \scriptsize 100.0]} \\
\bottomrule
\end{tabular*}
\end{table}

\begin{table}[t]
\footnotesize
\centering
\caption{Results and runtime from the trial with the lowest UC among five trials for each method in the Miller and Class-AB Op-Amp settings shown. \cRed{Red} and \cBlue{blue} fonts denote the best and second-best values for UC and runtime, respectively.}
\label{tab:ab1}

\setlength{\tabcolsep}{2pt} 
\renewcommand{\arraystretch}{1.1} 
\footnotesize

\begin{subtable}{\linewidth}
    \centering
    \caption{Results for the Miller Op-Amp.}

    \begin{tabular*}{\linewidth}{@{\extracolsep{\fill}}cccccccccc@{}}
    \toprule
    \textbf{PTM} 
    & \textbf{Run}   
    & \makecell{\textbf{Gain}\\($\geq$54dB)}     
    & \makecell{\textbf{PM}\\($\geq$60$^\circ$)} 
    & \makecell{\textbf{UGBW}\\($\geq$1MHz)} 
    & \makecell{\textbf{THD}\\($\leq$0.10\%)}   
    & \makecell{\textbf{Offset}\\($\leq$5mV)}   
    & \makecell{\textbf{Power}\\($\leq$0.15mW)}
    & \makecell{\textbf{UC}} & \makecell{\textbf{Runtime}\\(s)} \\
    \midrule

    \multirow{3}{*}{45nm}
     & VLM-CAD & 53.180 & 48.703 & 1.523  & 0.01269 & 3.838 & 68.321  & \cRed{2.405} & \cRed{2432.52} \\
     & Ablation 1 & 45.491 & 179.214 & 5.960 & 0.01460 & 3.529 & 111.095 & 9.250 & 2826.00 \\
     & Ablation 2 & 52.783 & 17.477 & 10.538 & 0.01425 & 2.524 & 115.607 & \cBlue{6.240} & \cBlue{2500.28} \\
    \bottomrule
    \end{tabular*}
\end{subtable}

\vspace{0.25cm} 

\begin{subtable}{\linewidth}
    \centering
    \caption{Results for the Class-AB Op-Amp.}
    
    \begin{tabular*}{\linewidth}{@{\extracolsep{\fill}}cccccccccc@{}}
    \toprule
    \textbf{PTM} 
    & \textbf{Run}   
    & \makecell{\textbf{Gain}\\($\geq$65dB)}     
    & \makecell{\textbf{PM}\\($\geq$50$^\circ$)} 
    & \makecell{\textbf{UGBW}\\($\geq$10MHz)} 
    & \makecell{\textbf{THD}\\($\leq$5.01\%)}   
    & \makecell{\textbf{Offset}\\($\leq$1mV)}   
    & \makecell{\textbf{Power}\\($\leq$10mW)} 
    & \makecell{\textbf{UC}} & \makecell{\textbf{Runtime}\\(s)} \\
    \midrule

    \multirow{3}{*}{180nm}
     & VLM-CAD & 85.505 & 51.897 & 35.111 & 0.001911  & 0.551 & 0.810 & \cRed{0.081} & \cRed{389.80} \\
     & Ablation 1 & 67.447 & 111.363 & 73.455 & 0.009684 & 0.138 & 3.120 & \cBlue{0.312} & 1437.40 \\
     & Ablation 2 & 90.765 & 52.568 & 13.447 & 0.00001075 & 0.018 & 3.176 & 0.318 & \cBlue{820.34} \\
    \bottomrule
    \end{tabular*}
\end{subtable}

\end{table}

We report the per-task Pass@1/Pass@3 and pooled statistics in Table~\ref{tab:sizing_results}. VLM-CAD improves pooled Strict Pass@1 over EESizer from 3.3\% [0.0, 8.3] to 23.3\% [10.0, 36.7] and Relaxed Pass@1 from 15.0\% [0.0, 36.7] to 91.7\% [80.0, 100.0]. Pass@3 moves in step, rising from 10.0\% and 16.7\% to 46.7\% and 99.2\% under the Strict and Relaxed modes, respectively, and VLM-CAD matches or exceeds EESizer on every task. For the VLM-CAD trials selected by UC, the corresponding total runtimes were 2432.52 seconds for the Miller Op-Amp and 389.80 seconds for the Class-AB Op-Amp, with Phase E accounting for 87.19\% and 40.81\% of these totals, respectively.

\subsection{Ablation Study}

To assess the significance of circuit schematics and the assistance of Image2Net, we conduct two ablations using Gemma-4-31B-it: 1) input only the original schematic, without Image2Net's annotation and JSON description; 2) input no schematic information at all. For each paradigm, we run five trials. Table~\ref{tab:ab1} reports the results: on the Miller Op-Amp, Ablation 1 performs worse and takes longer than Ablation 2, and both ablations yield higher and more unstable universal costs than VLM-CAD; on the Class-AB Op-Amp, both ablations meet the specifications, and VLM-CAD generally needs less total optimization time than the schematic-only ablation, while the no-schematic ablation can be competitive on easier settings.

\begin{figure}
\centering
\begin{minipage}{0.47\linewidth}
    \centering
    \includegraphics[width=\linewidth]{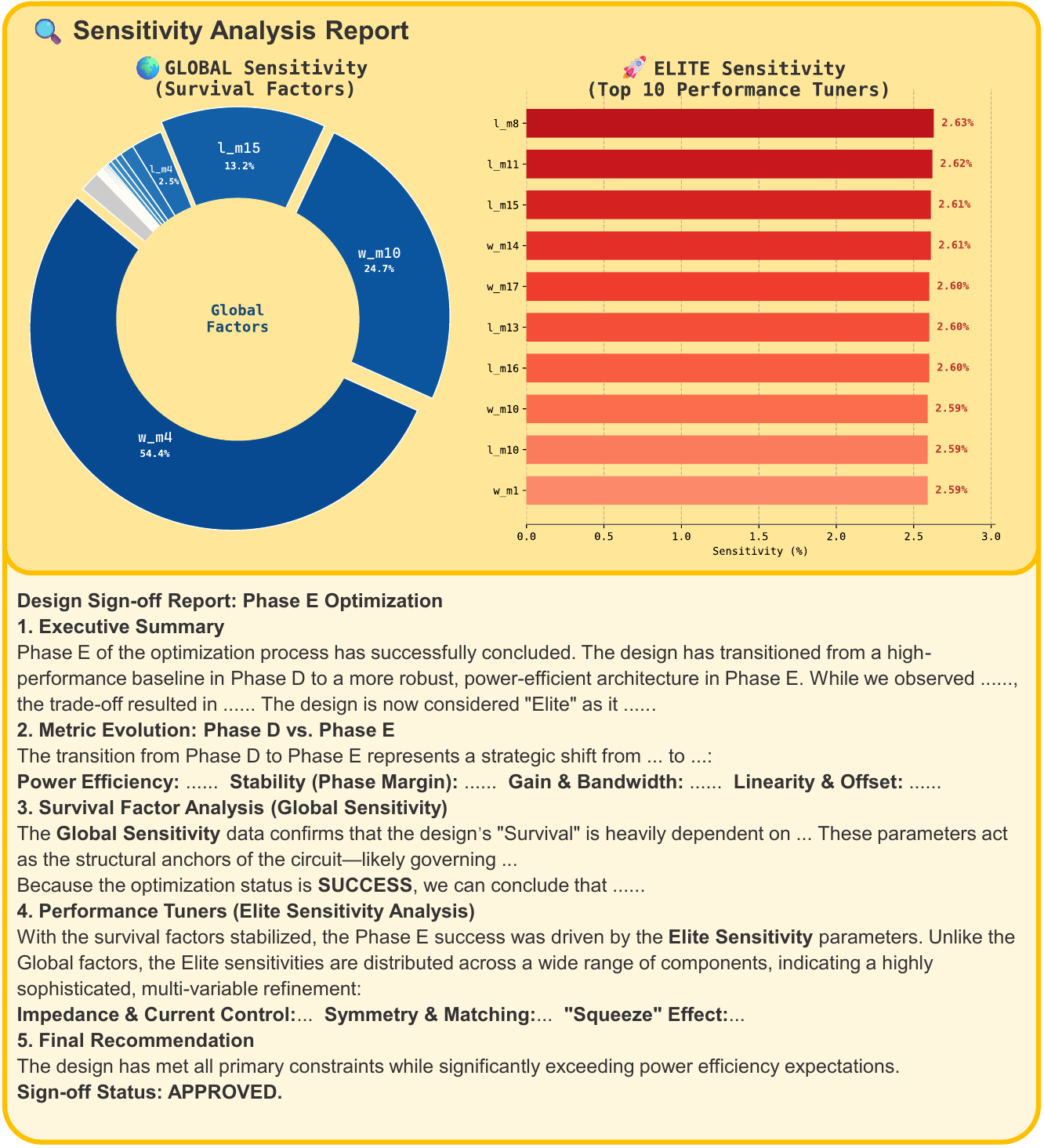}
    \captionof{figure}{Final design report including Global Sensitivity, Elite Sensitivity, and explanations.}
    \Description{Final design report including Global Sensitivity, Elite Sensitivity, and explanations.}
    \label{fig:report}
\end{minipage}%
\hfill%
\begin{minipage}{0.47\linewidth}
    \centering
    \includegraphics[width=\linewidth]{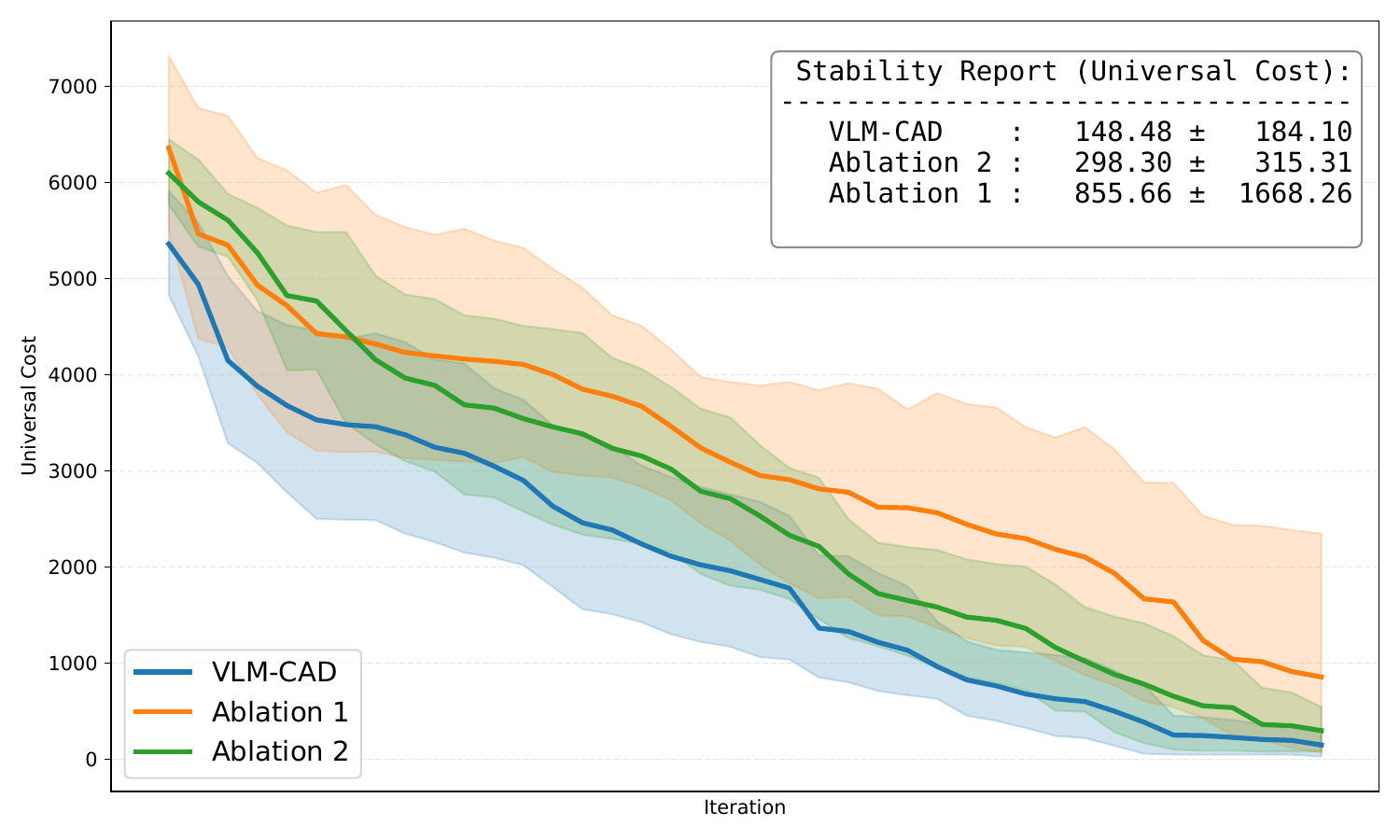}
    \captionof{figure}{Universal cost distribution of Phase D for the Miller Op-Amp. Both ablations show higher and more unstable costs than VLM-CAD, and ablation 1 performs worse than ablation 2.}
    \Description{Universal cost distribution of Phase D for the Miller Op-Amp. Both ablations show higher and more unstable costs than VLM-CAD, and ablation 1 performs worse than ablation 2.}
    \label{fig:cost}
\end{minipage}
\end{figure}

Ablation 1 being worse than Ablation 2 is consistent with a negative-transfer effect: the dense, repetitive schematic may act as a naturally adversarial visual input, consistent with VLM vulnerability to image-space perturbations~\cite{Cui_2024_CVPR} (see Figure~\ref{fig:cost}).

\subsection{Failure Analysis and Limitations}
Strict Pass@1 is zero on six of the twelve tasks. The per-run metrics reveal distinct failure signatures rather than a single mode: bandwidth limits the Discrete Op-Amp and the Class-AB Op-Amp, whose UGBW falls just below target while every other metric passes; gain limits DFCFC by a wide margin; and instability in the form of negative phase margin dominates NMCNR and PFC. The Miller Op-Amp tasks on 45nm BSIM PTM are the only ones that combine a marginal gain miss with a phase-margin collapse under the aggressive 18 MHz UGBW target, and Miller Op-Amp SpecsB is the sole task whose Relaxed Pass@1 falls to 40\%. Plausible contributors are the narrow feasible region imposed by coupled gain, bandwidth, stability, and power constraints, short-channel effects that weaken long-channel sizing heuristics, and warm-started trust regions entering locally attractive but infeasible basins. We treat these as hypotheses rather than causal findings.

Image2Net also remains imperfect: with a connectivity score of $0.8121\pm0.2116$, parsing errors can propagate into Phase B reasoning and later initialization. Finally, although the benchmark spans four technology platforms, it concentrates on analog amplifier topologies, so generalization to ADCs, PLLs, and power-management blocks remains open.

\section{Conclusion}
We propose VLM-CAD, a VLM-optimized collaborative agent workflow for analog circuit sizing that combines deterministic schematic grounding, agentic reasoning, and explainable trust-region optimization. Across 12 sizing tasks covering six circuits and four technology platforms, VLM-CAD improves pooled Strict Pass@1 from 3.3\% to 23.3\% over EESizer, while achieving 91.7\% Relaxed Pass@1. The structural evaluation shows that Image2Net provides stronger connectivity reconstruction than the evaluated general-purpose VLMs, and end-to-end runtime remains dominated by the simulation-driven ExTuRBO stage.

\begin{ack}
This research was supported by National Natural Science Foundation of China under Grant No.~62571173, Zhejiang Provincial Natural Science Foundation of China under Grant No.~LD25F020005 and No.~LQN25F030009.
\end{ack}

\bibliographystyle{unsrtnat}
\bibliography{VLM-CAD_ref}

\appendix
\section{Appendix: Experimental and Implementation Details}

\subsection{Experimental Environment}
Simulations use Ngspice-46 with DC operating-point, AC, and transient analyses. Experiments run on an AMD EPYC 7513 server with 64 cores, 256\,GB RAM, and four NVIDIA RTX 4090 GPUs. We serve the Gemma-4-31B-it backbone locally with vLLM, using tensor parallelism 4, a maximum model length of 126,000, the Gemma 4 tool-call and reasoning parsers, and GPU memory utilization of 0.96.

\subsection{Baseline and Execution Protocol}
VLM-CAD and EESizer operate on the same netlists, technology models, and specifications. EESizer runs up to 25 optimization iterations per run; the VLM-CAD DC-OP and inference-only phases run up to 10 and 40 iterations, respectively.

\subsection{VLM-CAD ExTuRBO Configuration}
ExTuRBO runs up to three parallel workers, each seeded from a distinct Phase-D candidate, with a 20-point initial cloud, a default local span of 40\% around the seed, and a 400-simulation budget per worker. For length parameters, the local bounds span 0.8 to 3.0 times the seed value. The Universal Cost uses penalty weights 1.0 for gain, 0.1 for phase margin, 0.2 for UGBW, 0.5 for THD, and 10.0 for offset, plus a sanity penalty of 100.0. Penalty terms act on raw engineering units, so the weights express relative emphasis rather than normalized margins.

\subsection{Statistical Detail}
\label{app:structural_protocol}
For structural evaluation, functional accuracy is computed as $F=100N_{\mathrm{correct}}/122$, where a prediction is correct when its circuit-type label matches the reference annotation. The connectivity score is computed per schematic as $S_k=1/\log_{10}(10+K)$, where $K$ is the graph edit cost between the predicted and reference connectivity graphs; the table reports the mean and population standard deviation of $S_k$ across the 122 schematics. A missing report entry is assigned $F=0$ and $K=|V|+|E|$ for the reference graph. An empty or unparsable predicted netlist receives the same $K$ penalty, while its $F$ remains determined by circuit-type label matching when available. The 2,268 schematics used to train Image2Net and the 122 noisy evaluation schematics contain disjoint circuit families.

Confidence intervals for pooled Pass@1 and Pass@3 are obtained by a task-level cluster bootstrap: we resample the twelve tasks with replacement 100{,}000 times with a fixed random seed, recompute the pooled mean per replicate, and take percentile intervals. Resampling at the task level preserves within-task dependence among the five runs.

\end{document}